\documentclass[preprint,12pt,3p]{elsarticle}
 \pdfoutput=1
\usepackage{mathptmx}
\usepackage[utf8]{inputenc}
\usepackage[numbers]{natbib}
 
\usepackage{graphicx}
\usepackage{float}
\usepackage[format=plain,
labelfont={bf,it},
textfont=it]{caption}
\usepackage{nth}
\usepackage{enumitem}
\usepackage{mathtools}
\usepackage{amsmath}
\usepackage{caption}
\usepackage{subcaption}
\usepackage{soul}
\usepackage{xcolor}
\usepackage{hyperref}
\usepackage{lineno, blindtext}
\DeclareMathAlphabet{\mathcal}{OMS}{cmsy}{m}{n}
\SetMathAlphabet{\mathcal}{bold}{OMS}{cmsy}{b}{n}

\begin{document}

\begin{frontmatter}

\title{The effect of surface morphology on the rate of phase change of micron and sub-micron sized 2-D droplets}

\author[label1]{Mohammad Rezaeimoghaddam}
\address[label1]{Department of Mechanical Engineering, Middle East Technical University, 06800 Çankaya, Ankara, Turkey}

\cortext[cor1]{Corresponding author}


\author[label1]{Zafer Dursunkaya}
\ead{refaz@metu.edu.tr}


%
%
%
\begin{abstract}
Heat transfer {\it via} phase change is a major contributor to heat removal in numerous engineering applications. Thin films of liquid result in increased heat transfer due to a reduction of conduction resistance, in addition the pressure jump at the liquid-vapor interface also affects the rate and direction of rate of phase change. Because of these effects the morphology of the substrate surface is expected to affect the film shape, hence heat transfer, especially in thin films. In this study, the influence of surface characteristics on the rate of phase change 
from micron and sub-micron sized 2-D droplets---i.e. films extending to infinity---forming on a substrate are modeled. 
Surface film profiles are generated on both flat and non-flat surfaces, triangular or wavy in nature, and a kinetic model for quasi-equilibrium phase change is applied.  In case of wavy surfaces, the surface is assumed to be a 
harmonic wave with an amplitude equal to the surface roughness and a wavelength corresponding to 
values commonly encountered in applications. Due to the presence of intermolecular forces at the contact line, which render the solution of the augmented Young-Laplace equation stiff, an implicit scheme is employed for the numerical integration. To verify the method, the predictions of a molecular dynamics (MD) simulation of a nano sized droplet present on a V-grooved surface is compared to the continuum model. The augmented Young-Laplace equation is solved numerically along with a phase change model originating from kinetic theory to calculate the shape of the two-phase interface forming the droplet and study the effect of various parameters 
on the rate of phase change.   Results are obtained for 
droplets with 
liquid pressures higher and lower than that of vapor, resulting in opposite contribution to phase change due to the pressure jump at the interface. The results show that the heat transfer rate can be substantially altered due primarily to the combined effects of surface morphology and disjoining pressure. 
\end{abstract}


\end{frontmatter}
\section{Introduction}
Micro heat transfer devices are widely used in cooling of computers, micro electro-mechanical systems and many other engineering applications where phase change occurs on nano and micrometer sized liquid films or droplets. They enhance the heat transfer rate due to the high values of latent heat of phase change process particularly in the close proximity  contact line where three phase meet and the film is thin. The rate of phase change is a strong function of the temperature and pressure difference at the interface as well as  the liquid film thickness.  For the last 50 years many studies have been performed to understand the phase change phenomena in the contact line region \cite{Wayner1971,Potash1972, Wayner1976,Stephan1992, DasGupta1993, Shanahan2001, Gokhale2003a, Gokhale2003, Panchamgam2005, Wang2007, Dhavaleswarapu2012, Hanchak2014, Yan2013, Akkus2016,Akkus2017,Bc2018,Alijani2018}.

When the dimensions of a liquid film/droplet is reduced to the order of micrometers, the shape of the liquid-vapor interface is dominated by the capillary forces and the effect of gravity subsides.   In case of even thinner liquid thicknesses, in addition to capillary pressure, $P_c$, the shape of the droplets in the close proximity of the contact line (e.g. $<100\,\rm{nm}$) is strongly affected by disjoining pressure ($P_d$), which is due to the inter-molecular forces between solid and liquid molecules given by the augmented Young-Laplace equation \cite{Derjaguin1978, N.V.Cburaev1987}. 
The variation of disjoining pressure with liquid thickness on a solid surface can be locally attractive or repulsive  or both \cite{Yeh1999a}. In the absence of structural and electrostatic effects, the molecular contribution---i.e. the dispersion effect---is the only component of disjoing pressure, following an inverse power relation with the liquid thickness \cite{VictorM.StarovManuelG.Velarde2007}. 
Wayner and Coccio \cite{Wayner1971} proved that due to the effect of disjoining pressure, the heat transfer process in the interline region is more efficient  compared to   simple conduction in an evaporating liquid meniscus.  The shape of the bulk fluid is affected by disjoining pressure and the evaporation or condensation rates are not only a function of temperature jump, but also of the disjoining pressure \cite{alipour2019,Akdag2019}.
The tendency of a liquid to wet a solid is characterized by the apparent contact angle, $\alpha_\mathit{app}$, the angle of intersection of the bulk fluid meniscus and the solid in the macro region.  Numerically, the apparent contact angle can be defined at the location where the disjoining pressure is negligible compared to capillary pressure and radius of curvature approaches a constant value \cite{Stephan1992}. 


Studies show that differences in surface morphology may result in significant changes to wettability in the contact line region and consequently change the rate of heat transfer \cite{Plawsky2009, Ojha2010, M.Ferrari,Plawsky2014}. 
When smooth surfaces commonly used in engineering applications are observed, the ratio of the surface roughness  heights to the distance between two consecutive surface peaks ranges between $1/10$ to $1/100$ with a typical value of $1/20$. The average value of surface roughness is in the range of $0.05$--$0.8\mu{\rm m}$ and the average distance of crest spacing or wavelength---i.e mean peak spacing---is reported in the range of $2$ -- $40\mu{\rm m}$ in the literature \cite{Ryan200, H.M.SIDKI2001, Meli2002, Taylor2006, Young2009, Adams2012, Jouini2009, Young2007, Wen2006, Poon1995, Montero-Moreno2007, Rauf2009, kubiak2011}. Therefore, surface film profiles can be assumed to be a harmonic wave with an amplitude equal to the surface roughness $\,\epsilon$, and a wavelength $\lambda$, with a surface roughness to wavelength ratio, $\epsilon/\lambda$, commonly encountered in applications. 
It is expected that quantifying the liquid shape in a 2-D film or a 3-D droplet, as well as contact angle in the micro region is crucial in understanding the operation of micro heat transfer devices. However, near the contact line, continuum mechanics breaks down due to extreme thin liquid films, requiring the use of molecular interactions to assess the micro contact angle, $\alpha_\mathit{micro}$, at the intersection of the solid-liquid-vapor phases. The micro contact angle is defined as the angle formed by a line drawn through the centers of atoms lying in successive molecular layers at the edge of the droplet (or film) with respect to the horizontal \cite{Yeh1999}. 
In the case of a bulk of fluid filling a micro/nano-structured surface, geometry  has a deciding effect on the 2-D film (or droplet) shape whether it is convex, concave or flat.
 

A model of capillary condensation on equilibrium menisci of wetting films on triangular rough solid surfaces was used and the augmented Young-Laplace equation was solved by using Galerkin/finite element method \cite{Sweeney1993}. 
The model used did not cover all the possible ranges of surface roughness used in real engineering devices and also did not simulate convex droplet shapes. Recently, the microscale transport processes that occur in the contact line region of a liquid corner meniscus for four different silica surfaces of various surface roughness were studied \cite{Ojha2010}.  It was found that that the evaporative 
heat flux was higher on the rough surface and as a result, the rough surface had higher rates of evaporative heat transfer compared to a flat substrate due to the increase in surface area. 

Plawsky et al. \cite{Plawsky2009, Plawsky2014} reviewed the  effects of surface topography, nano scale surface modification, surface chemistry and fluid physics on evaporation at the contact line. They suggested
 that altering the surface chemistry and surface topography on the micro- and nanoscales can be used to drastically enhance vaporization. 

  Molecular dynamics (MD) simulation is an alternative approach commonly employed in the study of 2-D and 3-D spherical cap droplets on a flat surface \cite{Ji2008,Maroo2011,Hens2014,Zhang2015}, however, computationally expensive. The formation of thin films on a nano structured surface was simulated with MD for different fluid-solid systems in numerous studies \cite{Hu2014,Kim2016,Niu2018}; however, due to limitations in the temporal length scales in MD simulations, comparison of MD simulation results with continuum models using augmented Young-Laplace equation is not possible for large droplets. MD in conjunction with volume of fluid (VOF) method was used to study a multi-scale simulation of dynamic wetting of water droplets spreading on a platinum surface \cite{Zhang2017}. 
  The system did not cover large droplets to compare the results for meso-scale and micro sized droplets.
  
%

Both continuum models and MD simulations have inherent shortcomings.  In continuum based models numerical accuracy and stability of simulation at contact line region requires special attention, in addition to the absence of knowledge of the value of the micro contact angle, which is required to serve as a boundary condition at the contact line. The MD simulations, on the other hand model the interface shape, and hence predict the micro contact angle only for small systems, due to time scale and system size restrictions, rendering them unpractical for micron size or larger droplets/films.  In the current study, it is proposed to use a hybrid approach, where MD is used to calculate the micro contact angle for nano to micro sized droplets, extrapolated to the limit of continuum, beyond which the macro solution is obtained using a continuum based model of augmented Young-Laplace equation on the interface.  In this approach, the value of micro contact angle obtained from MD simulations is used as a boundary condition for augmented Young-Laplace equation solution. 


In this study, the influence of surface characteristics on phase change rate of a fluid partially filling a non-flat surface is investigated. The thin liquid film is a 2-D nominally cylindrical droplet, in effect a liquid film.  Due to the small scale, the effect of gravity is absent and the disjoining pressure important, and in some cases dominant.
To the best knowledge of the authors, effects of real surface morphology on the rate of heat transfer for micro sized droplets are missing in the literature. In such a problem initially the shape of the droplet liquid-vapor interface should be found. Thus the main objective is to obtain an accurate 2-D droplet profile especially in the contact line region and then apply the heat transfer model based on the kinetic theory. The values of micro contact angle and liquid film thickness in the proximity of the substrate are unknown and need to be determined accurately to obtain reliable results. Moreover, the maximum heat transfer rate occurs in the contact line region and the accuracy and robustness of the solution of augmented Young-Laplace equation is important. For this problem the liquid is exposed to its vapor and an interface separates the two phases.  The surface is characterized by a harmonic function whose dimensions correspond to surfaces encountered in common traditional and nano manufacturing processes. The shape of the quasi-static surface is generated by solving the force balance due to the capillary and dispersion effects, and phase change rate calculated using expressions given by kinetic theory. 
\section{Modeling and Solution Approach}

The problem studied covers a range of two-dimensional droplets with the largest radius of curvature of the order of micrometers, and the smallest approaching the limit of continuum.  For length scales of these magnitudes, the effect of gravity is negligible and the condition of static equilibrium is governed by the augmented Young-Laplace equation derived from classical thermodynamics \cite{Derjaguin1978,DeGennes1985,N.V.Cburaev1987,Brochard-Wyart1991,Yeh1999a,VictorM.StarovManuelG.Velarde2007},
\begin{equation}
\label{eq:AYLE}
P_v-P_l=P_c+P_d\equiv \Delta P_\mathit{int},
\end{equation}
where $P_v-P_l$, the pressure difference between vapor and liquid phase of working fluid, or the pressure jump at the interface, $\Delta P_\mathit{int}$, determines the droplet volume.  A smaller value of pressure difference corresponds to a larger droplet.  The capillary pressure $P_c$ for a 2-D cylindrical surface is given as,
\begin{equation}
\label{eq:Pc}
P_c = \frac{\sigma}{R} = \frac{\sigma \delta_{xx}}{(1+\delta_x^2)^{3/2}},
\end{equation}
where $R$ is the radius of curvature of the liquid-vapor interface, $\sigma$ the surface tension and the subscript $x$ denotes differentiation along the coordinate axis.  For planar films separating semi-infinite solids, the disjoining pressure, $P_d$, is a function of the film thickness, $\delta$,
\begin{equation}
\label{eq:Pd}
P_d=\frac{A}{6\pi \delta^3},
\end{equation}
where $A$, is the Hamaker constant and $\delta$ is the thickness of the liquid film. 
The film thickness $\delta$ is defined as the shortest distance from the surface of the liquid to the substrate \cite{Hu2014}. 
Rearranging \autoref{eq:AYLE}, combining the definitions of capillary and disjoining pressures given in \autoref{eq:Pc} and \autoref{eq:Pd} and writing in terms  of $\delta_{xx}$, gives
\begin{equation}
\label{eq:delta_dp}
\delta_{xx}= \bigl( 1+\delta_x^2 \bigr)^{3/2} \ \Bigl(\frac{A}{6\pi\delta^3}-\Delta P_\mathit{int} \Bigr).
\end{equation}
To generate the shape of the equilibrium interface, \autoref{eq:delta_dp}, a nonlinear ordinary differential equation for the film thickness must be solved numerically.  This can be done by converting \autoref{eq:delta_dp}  into a system of simultaneous first order ODEs, and subsequent numerical integration. 

The solution of the boundary value problem defined by \autoref{eq:delta_dp} requires two boundary conditions.  One is the derivative of film thickness, $\delta_x$, at the origin of integration and the second one is the symmetry condition at the line of symmetry.  
%
%
In the absence of disjoining pressure, the solution is one of constant radius of curvature---due to the negligible effect of gravity---and the slope at the contact line is set by the apparent contact angle. 
The pressure difference between the vapor and the liquid, $\Delta P_\mathit{int}$, sets the radius of curvature of the droplet, hence the cross sectional area of the droplet in the 2-D problem---or the volume in a 3-D problem.

\begin{figure}
\centering
\includegraphics[scale=0.5]{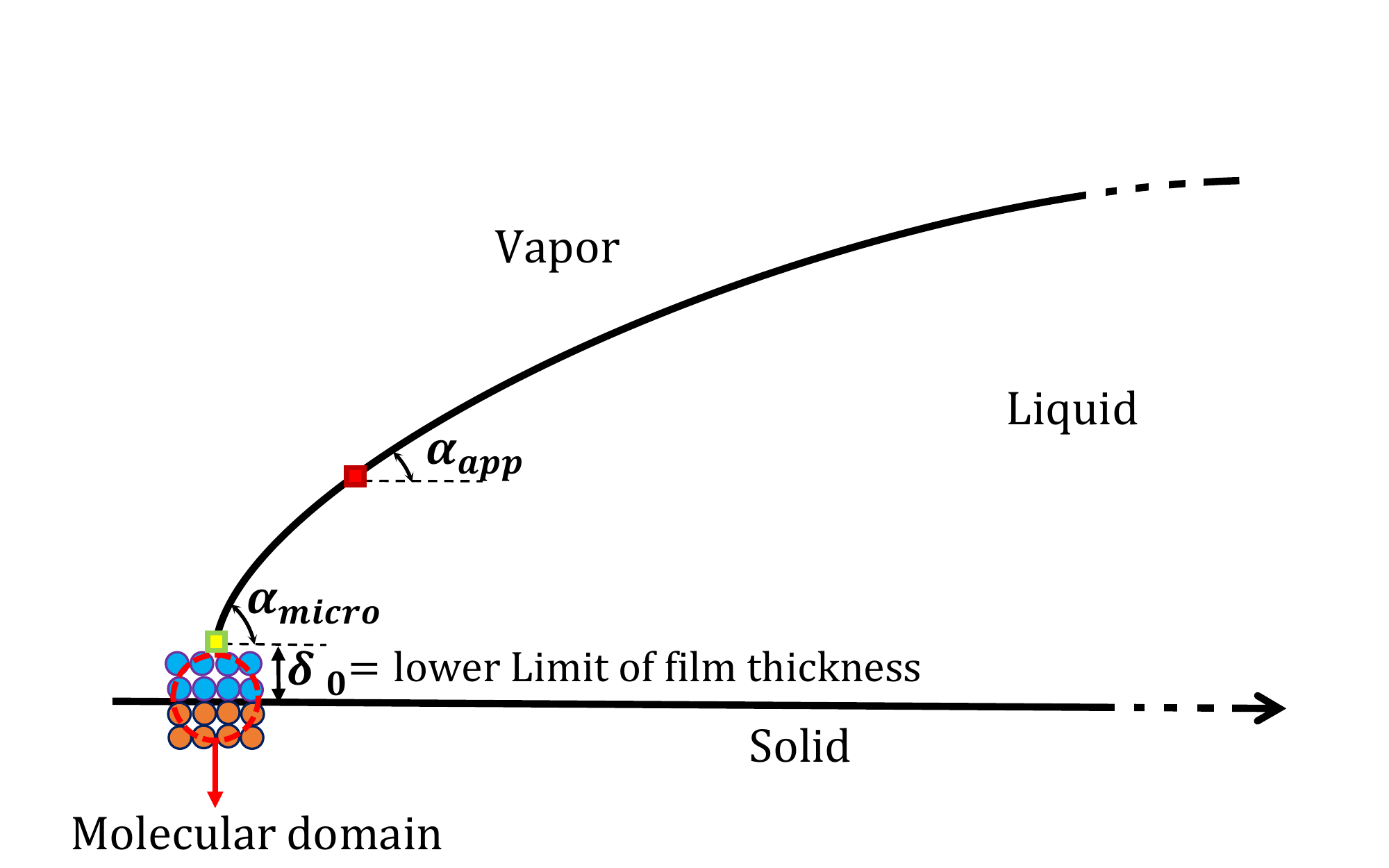}
\caption{Nanoscopic/microscopic zone near the static triple line of a liquid on a smooth solid partially adopted from \cite {DeGennes1990, Shanahan2001}}.
\label{FIG:numericfig}
\end{figure}

When the effect of disjoining pressure is included, the solution deviates from one with a constant radius of curvature, particularly in the close proximity of the solid substrate.  In this case the value of the derivative at the wall is also not set by the apparent contact angle, since the apparent contact angle is an experimentally determined quantity, measured on a scale larger than where dispersion effects are weak.  Furthermore, a molecular domain, a few Angstroms in thickness, exists in the close proximity of the contact line, where the continuum description breaks down \cite {DeGennes1990, Shanahan2001,Bonn2009}.  In the current study this is taken to be the lower limit of film thickness that can be modeled with the current continuum model.  \autoref{FIG:numericfig} shows the schematic view of nanoscopic and microscopic triple line of a bulk of liquid that lies on a smooth surface.  \autoref{FIG:numericfig} also indicates the point where the dispersion effects become negligible and the angle of the tangent to the interface approaches the apparent contact angle.   
In the presence of dispersion effects due to the deviation from a constant radius of curvature, the value of the micro contact angle at the wall, $\alpha_\mathit{wall}$, is different from the apparent contact angle, and is an unknown. Yeh et al. \cite{Yeh1999} stated that for macroscopic drops the macroscopic contact angle is obtained by an appropriate interpolation from the droplet shape in the constant curvature region.

Since in the presence of dispersion effects the value of the first derivative at the wall,  $ \delta_x|_{x=0} = \tan(\alpha_\mathit{wall})$, cannot be determined, to overcome this and integrate the differential equation, an iterative numerical solution approach is adapted, which is as follows:  starting from an initial fluid film thickness, $\delta_0=10{\rm \AA}$, excluding the molecular layer, \autoref{eq:delta_dp} is integrated along the surface using an initial estimate for the actual contact angle until a point on the interface where the disjoining pressure term is negligible compared to capillary pressure.  At this point the slope of the liquid film, $d\delta/ds$ is expected to be equal to $\tan(\alpha_\mathit{app})$.  The solution is repeated using shooting method in conjunction with a secant algorithm to obtain the value of the micro contact angle that results in a liquid slope giving the apparent contact angle at the location where the magnitude of disjoining pressure is negligible compared to the capillary pressure, $\bigl|{P_d}/{P_{c}}\bigr| < \varepsilon$, where a value of $0.001$ is used for $\varepsilon$ is $0.001$ in the current study.  The angle the liquid film makes with a hypothetical line parallel to the surface at this point, will be adapted as the definition of the apparent contact angle in the current study.

\section{Results and Discussion} 

\begin{figure*}
\centering
\begin{subfigure}{.5\textwidth}
  \centering
  \includegraphics[width=1.1\linewidth]{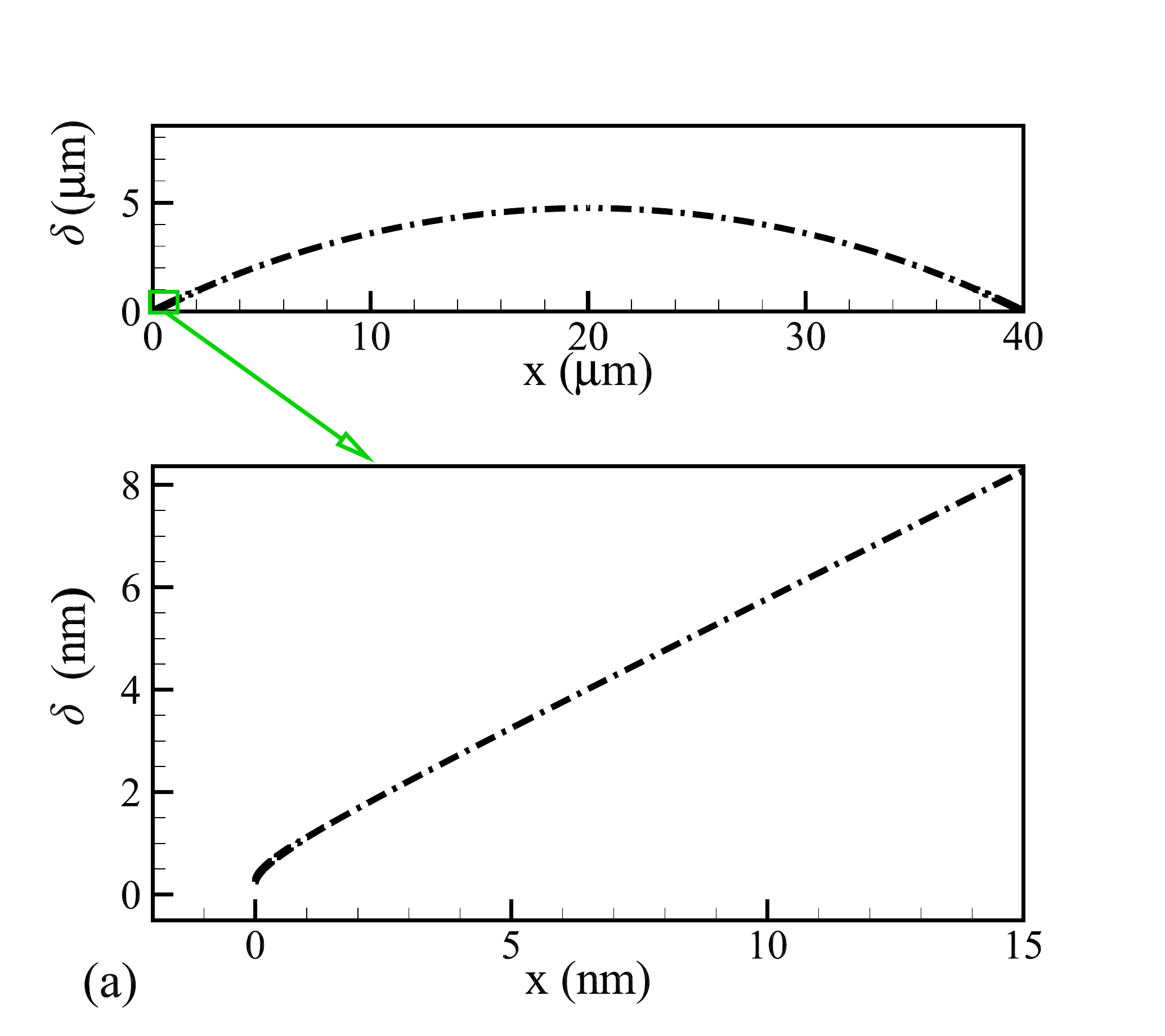}
\end{subfigure}%
\begin{subfigure}{.5\textwidth}
  \centering
\includegraphics[width=1.1\linewidth]{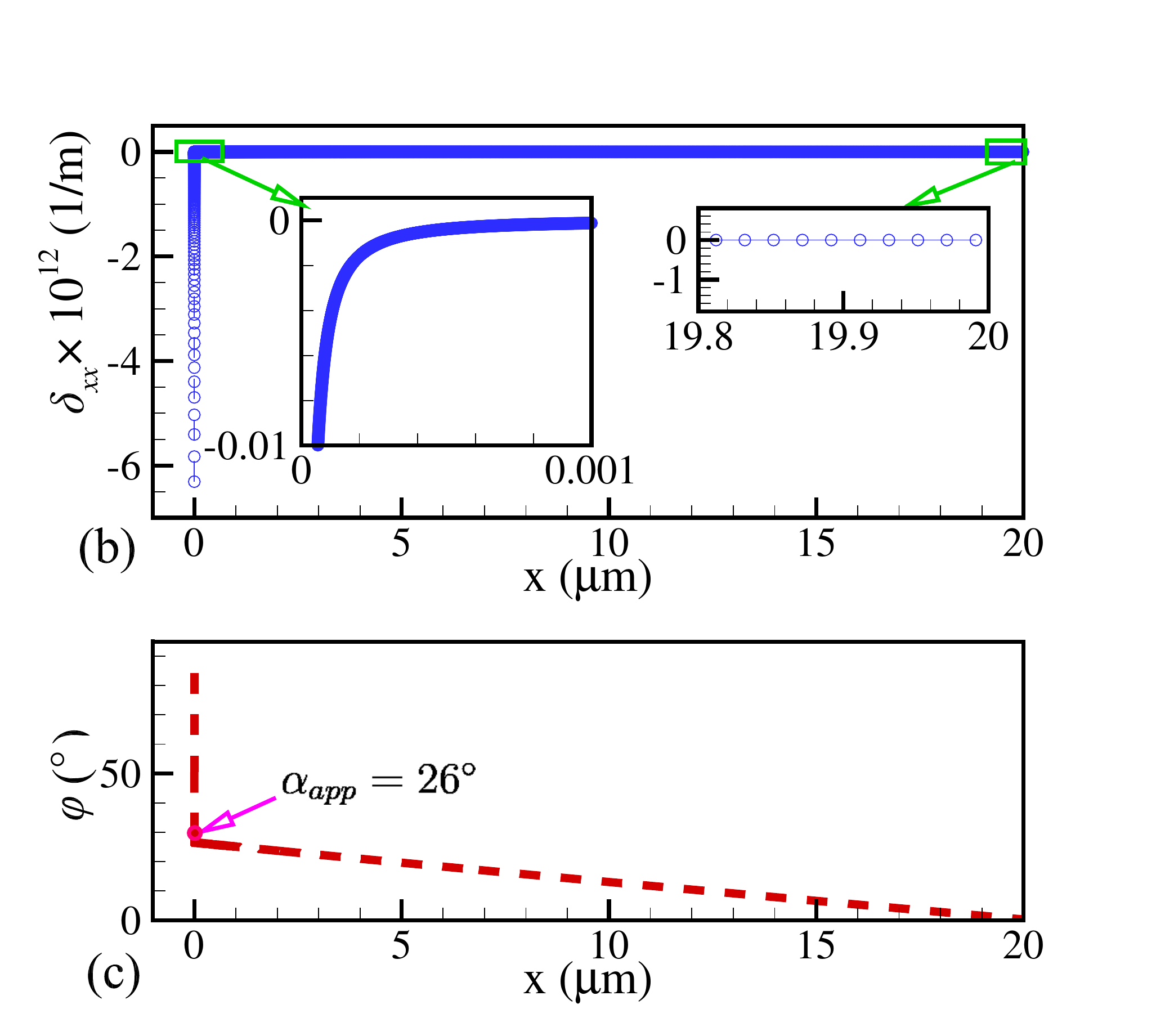}
\end{subfigure}
\caption{Plots of $\delta$, $\delta_{x}$ and $\delta_{xx}$ for a  droplet with apparent contact angle of $26^\circ$ (a)  variation of 2-D cylindrical droplet profile of length $40 \mu {\rm m} $ and its
close up view in the proximity of the surface (b) variation of the second derivative of film thickness, $\delta_{xx}$, and (c) variation of droplet surface angle along the length of droplet ($d\delta/dx \rightarrow \rm{\varphi}$).} 
\label{FIG:stiffness}	
\end{figure*}

The integration of \autoref{eq:delta_dp} is straightforward in the absence of dispersion term, i.e. when the Hamaker constant, $A$, is set to zero.  When the disjoining pressure is included, however, due to the highly nonlinear nature of the dependency on film thickness, $\sim 1/\delta^3$, which approaches infinity on the wall, the numerical problem is stiff for small values of the film thickness when $\delta < 100 {\rm nm}$, which renders explicit numerical methods---such as Runge-Kutta or predictor-corrector based schemes---unstable.  The simultaneous presence of rapidly changing components together with slowly changing ones necessitates the utilization of appropriate step sizes in different domains where the numerical solution is attempted.  This is accomplished by the implementation of an adaptive step size in the integration process which both reduces the computational effort and also simultaneously avoids introducing errors due to small increment of step-size and controls the accuracy while achieving robustness and stability of the solution \cite{Akkus2016}.  \autoref{FIG:stiffness} shows the variation of the second derivative, $\delta_{xx}$, in the domain, where the integration starts from an initial value of the film thickness $\delta_{0}=3 \AA$. 
In the close proximity of the wall, the magnitude of the disjoining pressure is very large, and to balance this term, the capillary pressure term is also large with the opposite sign resulting in second derivatives $\mathcal{O}(6.5\times 10^{12} {\rm /m}) $, which correspond to a radius of curvature of $ {0.16\rm nm}$, the magnitude of which rapidly decreases with increasing film thickness due to the fading dispersion effect.  When the effect of disjoining pressure subsides, the radius of curvature attains a constant value of approximately $2\times 10^{-4}{\rm /m} $ at the apex, with a negative second derivative, corresponding to a radius of curvature of $ 5.4 \mu{\rm m}$, staying constant till the end of integration at the line of symmetry.  Although the change in the second derivative appears to be discontinuous on a large scale, the close up view in the inset shows a resolved and smooth transition to a constant value.  The figure also indicates the step sizes taken during this computation by marking the individual points of solution, showing that the step size is relaxed by many orders of magnitude as the integration progresses away form the solid surface.  The ratio of the initial and final step sizes used in the simulation reaches approximately $1:1000$. 

\begin{figure}
\centering
\includegraphics[scale=0.5]{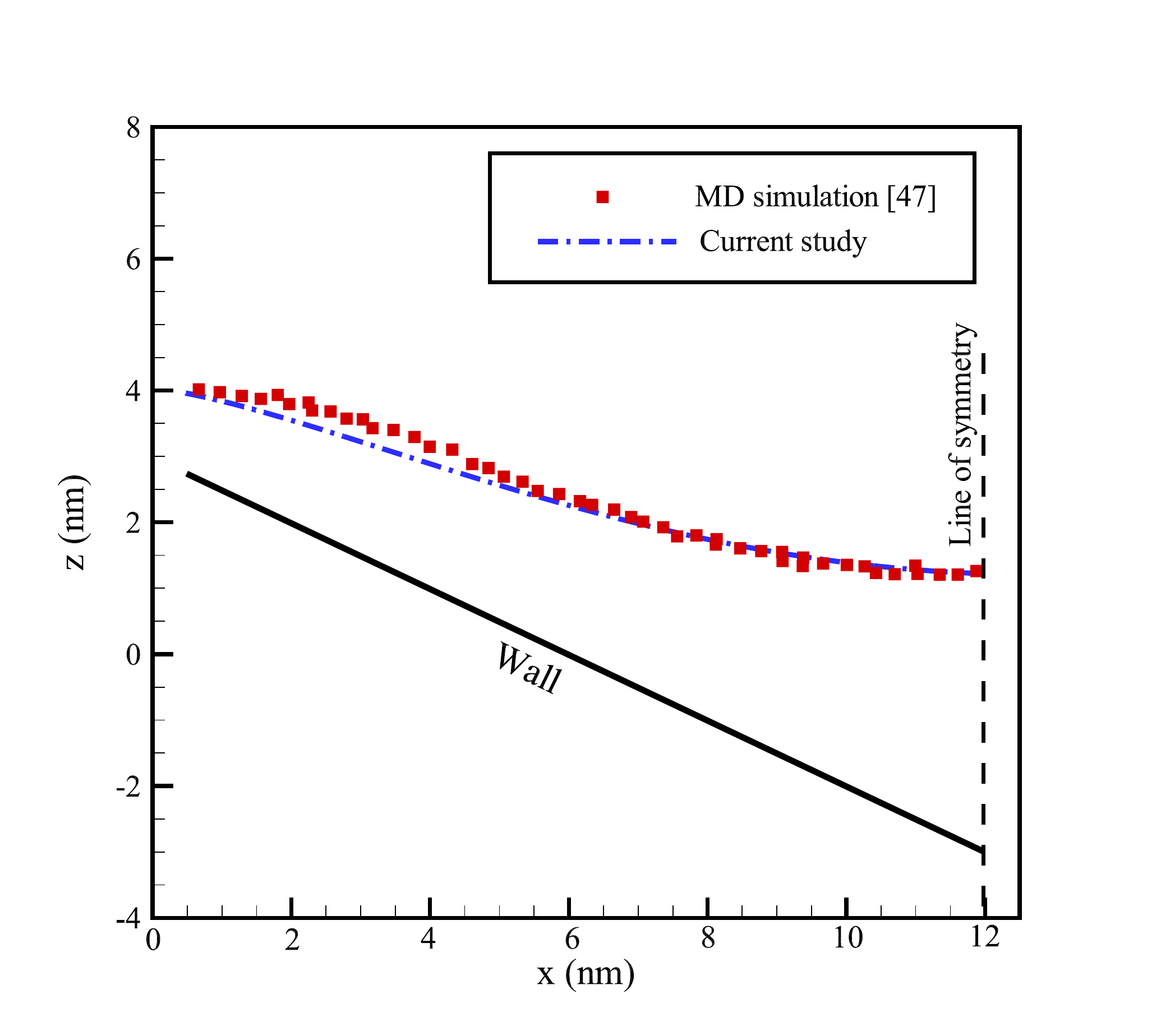}
\caption{Comparison of molecular dynamics data with the results of augmented Young-Laplace equation based model for a droplet on a V-grooved nano
structured surface.}
\label{FIG:validation}
\end{figure}

The first case to be simulated is one where the droplet is very small and the interface never satisfies the above defined condition of capillary pressure dominating the disjoining. 
As a result, the dispersion effects are always present and the surface is formed due to a balance between the two constituents of the interface force balance, both the disjoining and capillary pressures.  To solve this case the value of the contact angle at the wall is required.  A molecular dynamics simulation of  droplet formation on a nano-structured surface given in \cite{Hu2014} is used to assess the value of the actual contact angle at the wall. 
Here, the molecular dynamics simulation of
an initial cluster of particles forming a thin water film of thicknesses $2.48 {\rm nm}$ was placed on a nanostructure of depth $D = 5.71 {\rm nm}$,  \cite{Hu2014}, are compared to the current augmented Young-Laplace equation results.
In the solution to this initial value problem, the value of the film thickness at the beginning of the domain of integration is set to $\delta_0=1.4 {\rm nm}$  and the value of the first derivative, $\delta_x$, are both obtained from the results of a molecular dynamics simulation \cite{Hu2014}. 
This case is simulated using the current approach and the predictions of both simulations are given in \autoref{FIG:validation}. 
In this figure, the solid line indicates the solid surface, and the vertical coordinate axis, which show the fluid thickness is set at the center of the V-groove.  The result shows the agreement between the predictions of the molecular dynamics and current 
model, which in fact confirms that the functional dependence of disjoining pressure on film thickness given in \autoref{eq:Pd} in the continuum approach matches the Lennard-Jones potential based molecular dynamics model adequately.  In this simulation the value of Hamaker constant $A$ is $3\times 10^{-19}$ J, given in \cite{Hu2013}.

\begin{figure*}
\centering
\begin{subfigure}{.5\textwidth}
  \centering
  \includegraphics[width=1.1\linewidth]{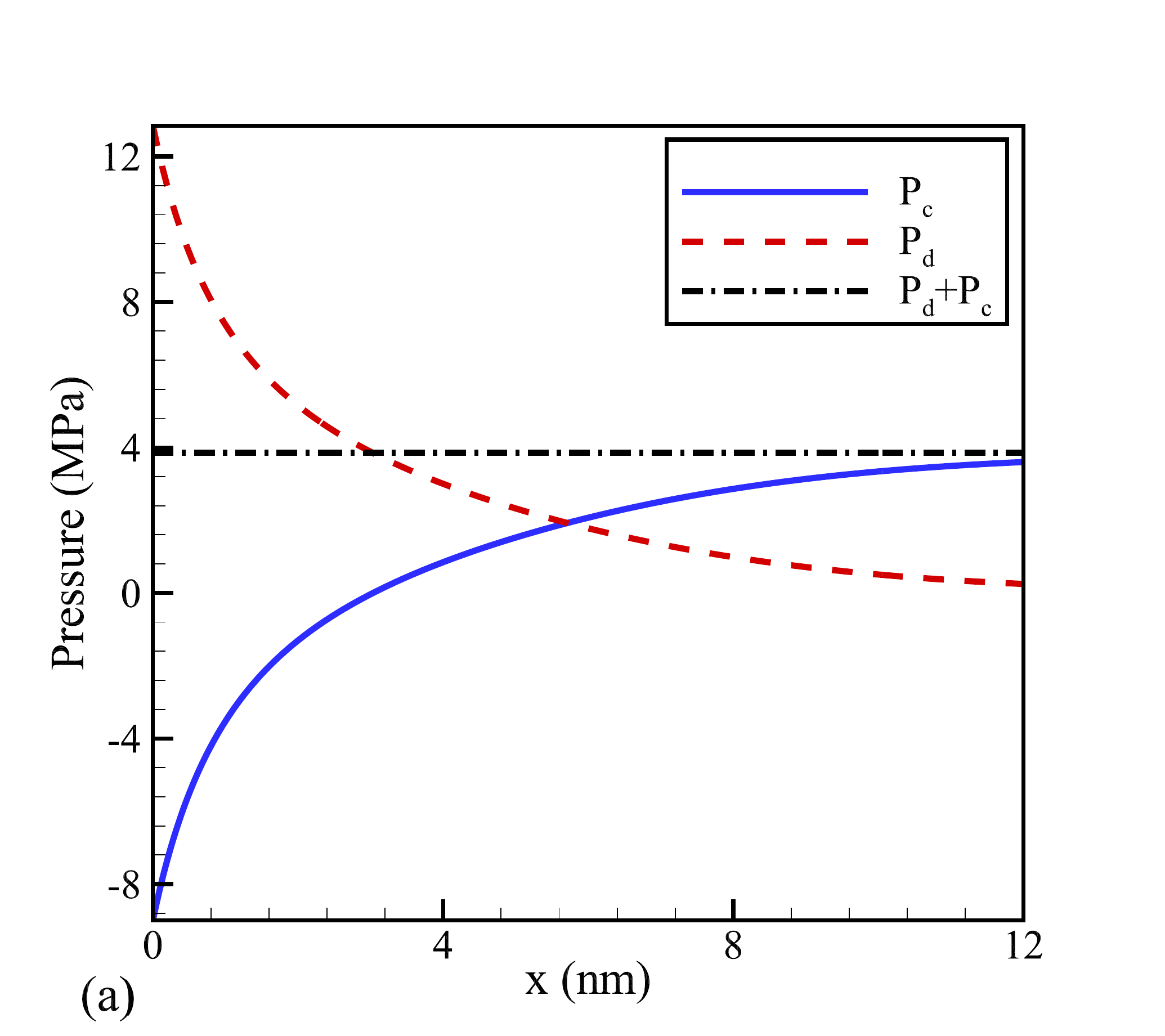}
\end{subfigure}%
\begin{subfigure}{.5\textwidth}
  \centering
\includegraphics[width=1.1\linewidth]{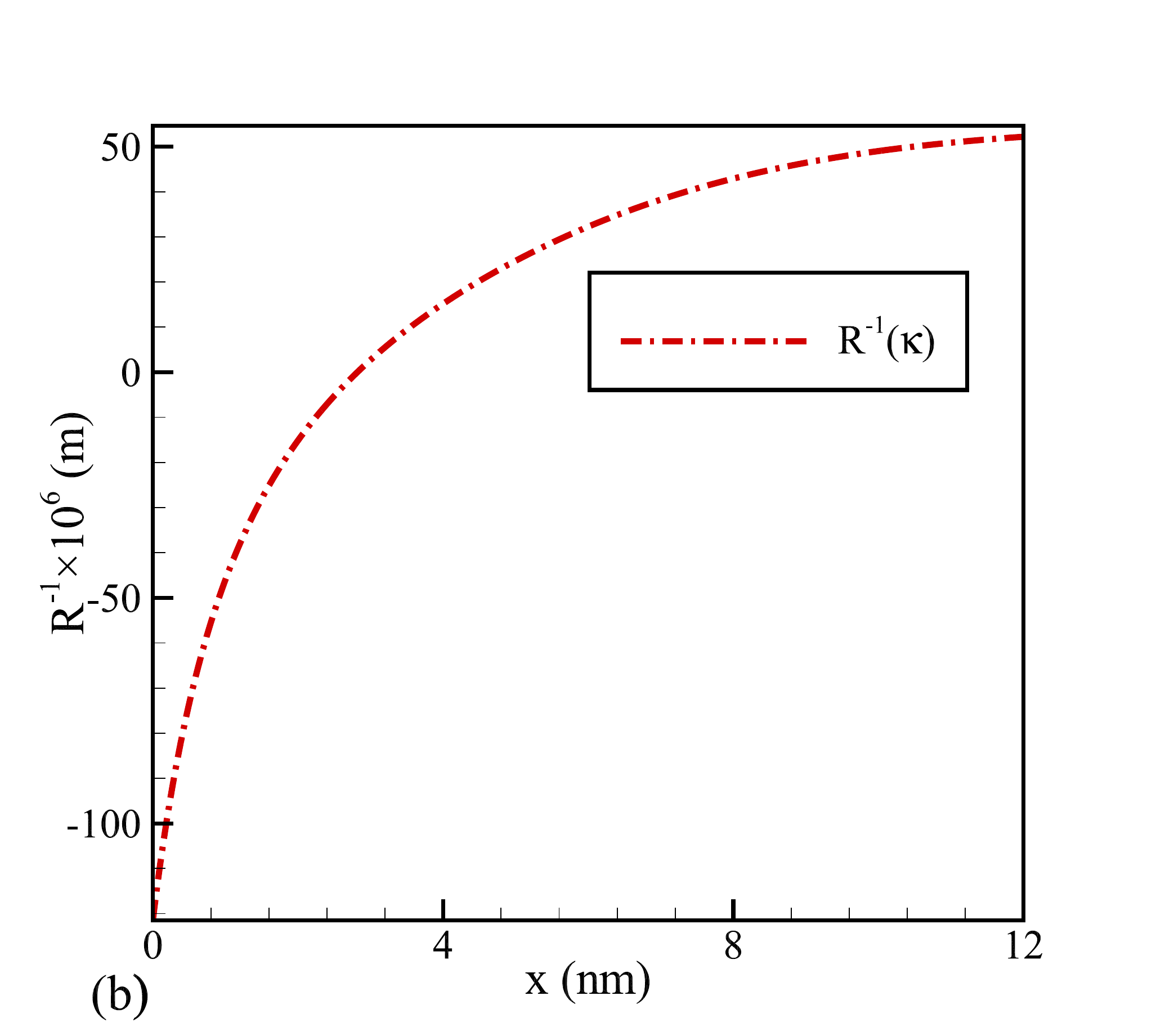}
\end{subfigure}
\caption{ 2-D droplet on a V-grooved nano structured surface variation of (a) pressure (b) curvature ($\kappa$) along the length of the droplet.}
\label{FIG:Pressure_valid}	
\end{figure*}

\begin{figure*}
\centering
\includegraphics[scale=0.55]{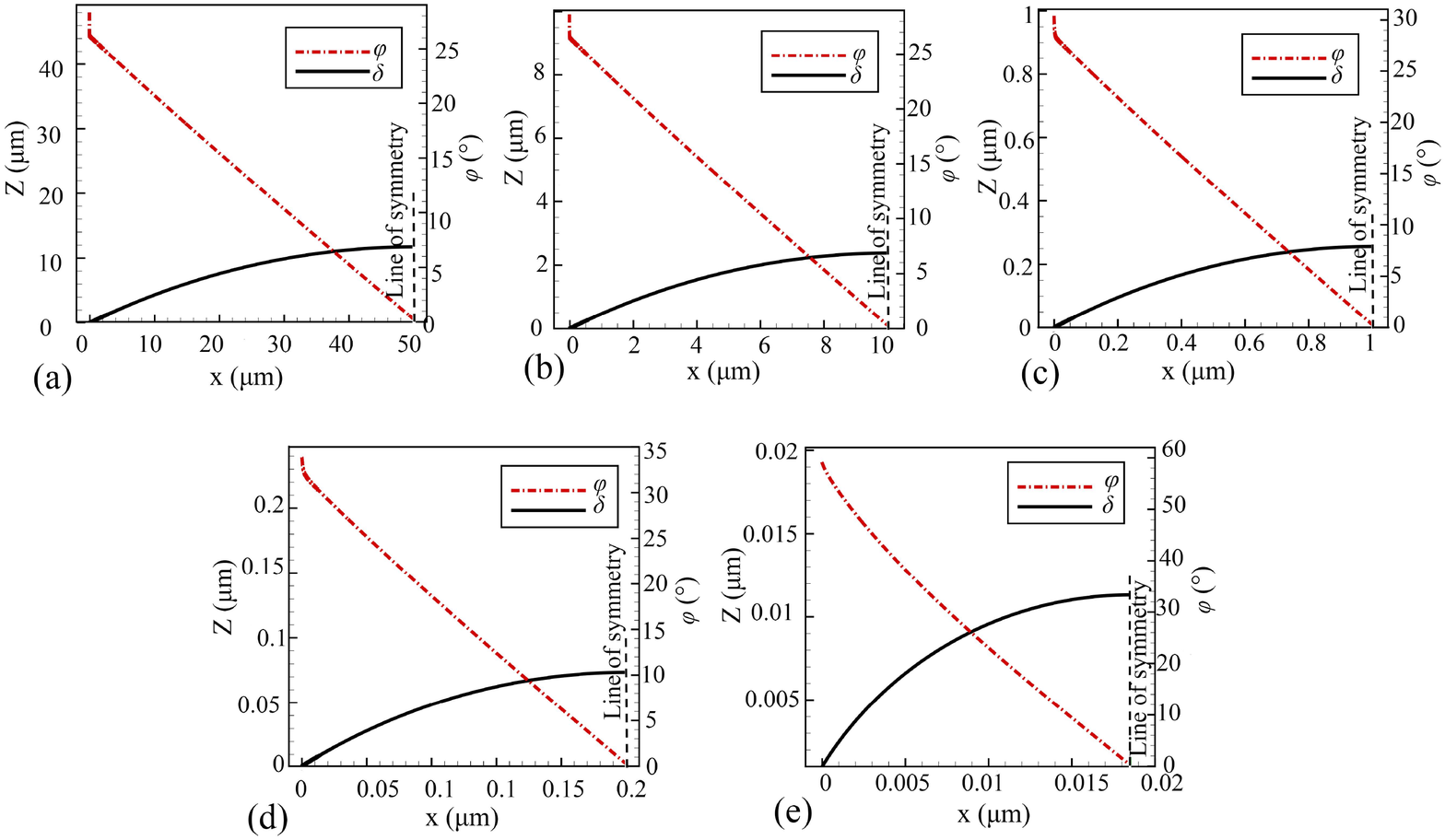}
\caption{Variation of droplet profiles with the same apparent contact angle ($\alpha_{app}=26^\circ$) along the surface. (a), (b), (c), (d) and (e) are 2-D cylindrical droplets with $100, 20, 2, 0.4$  and $0.037 \rm{\mu m}$ length, respectively. Plotted between the edge and line of symmetry.} 
\label{FIG:contactAnglescale}
\end{figure*}

\begin{figure}
\centering
\includegraphics[scale=0.4]{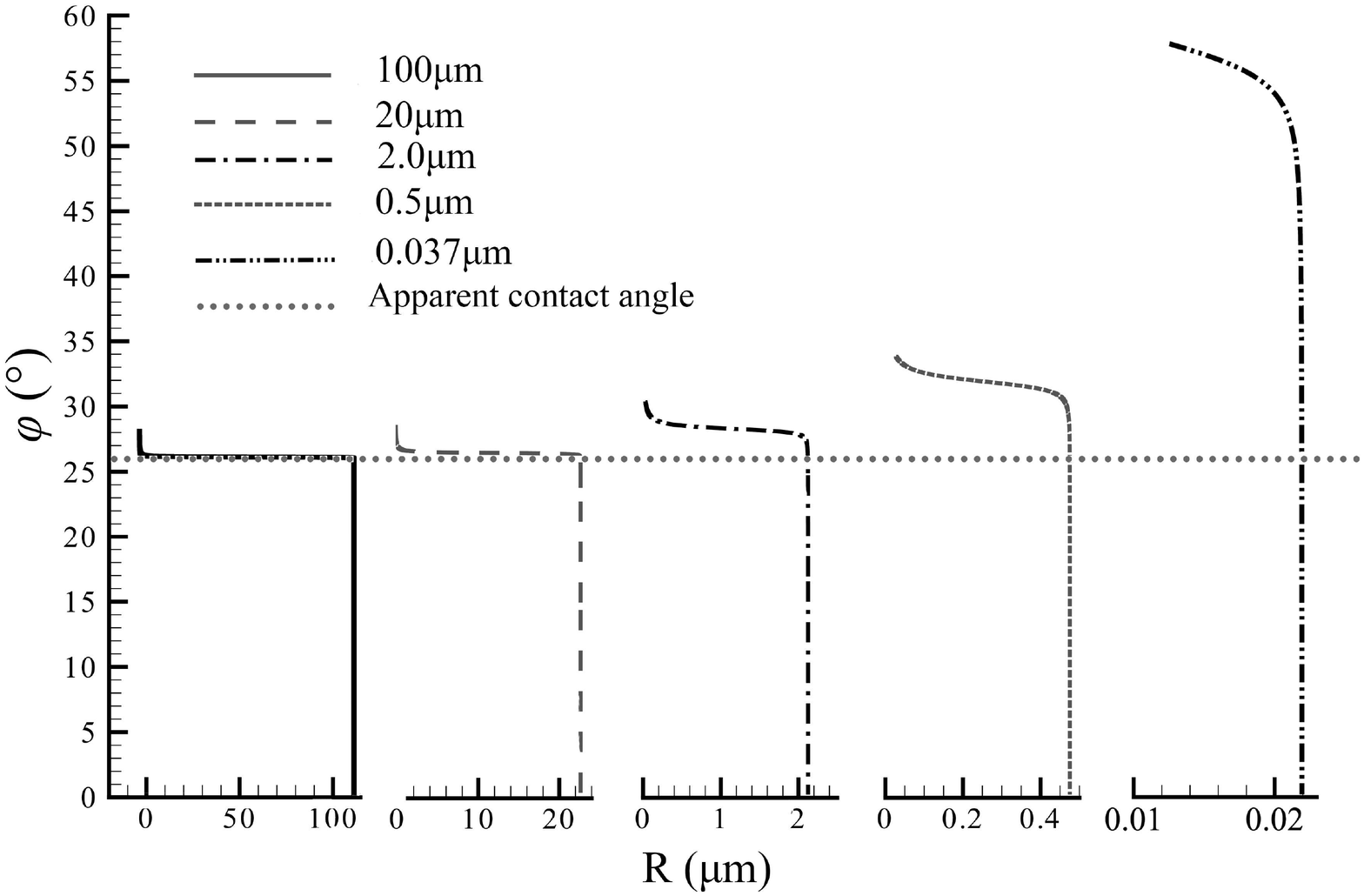}
\caption{Variation of droplet profile slope, $\varphi$, with radius of curvature for different size droplets and identical apparent contact angle ($\alpha_{app}=26^\circ$). }
\label{FIG:Angleradiusofcurvature}
\end{figure}

In this problem, due to the absence of gravity, the pressure difference between the liquid and vapor remains constant on the interface (\autoref{eq:AYLE}).  \autoref{FIG:Pressure_valid} shows the variation of disjoining and capillary pressures, which sum up to the constant pressure jump, $4 \,{\rm MPa}$, in this case. 
  The large magnitude of the pressure jump is due to the excessively small size of the droplet. Approaching the line of symmetry, which is  $12\,{\rm nm}$ away from the contact line, the magnitude of the disjoining pressure term decreases to almost zero, compensated by an equivalent increase in the capillary pressure.  The curvature, $\kappa$, which is the inverse of radius of curvature, starts with a negative value, indicating a concave surface and changes to positive away from the contact line, due to decreasing dispersion effect.  The curvature never attains a constant value, indicating the presence of disjoining pressure even at the point of thickest film at the line of symmetry.
%
%
%
%
The shape of the interface is generated for 2-D films with identical contact angles but each having a different film thickness at the line of symmetry, increasing with increasing size.  The study ranges between the smallest film where an apparent contact angle can be defined---i.e. where the disjoining pressure is negligible compared to the capillary in the neighborhood of the line of symmetry---approximately $37\ {\rm nm}$ wide for the current case, to a size where the effect of gravity becomes sensible, approximately $100{\rm\ \mu m}$ for this problem. The apparent contact angle, $\alpha$, is $26^\circ$, and the cylindrical droplets are $100, 20, 2, 0.5$ and $0.37\ \mu {\rm m}$ wide.  The variation of film thickness and the angle $\varphi$ that the surface tangent forms are given in \autoref{FIG:contactAnglescale}.
\autoref{FIG:Angleradiusofcurvature} shows the variation of the surface slope on the interface from the contact line to the line of symmetry for all 5 cases studied. 
 The angle $\varphi$ is a direct indication of the slope, given by $\delta_x = \tan\varphi $.  When the film width (droplet size) is large, the micro contact angle is close to the apparent one at the wall, reaching the apparent contact angle with a rapid change, at which point the change ceases due to the vanishing disjoining pressure, and the radius of curvature reaches a constant value thereafter.  In the case of small droplets, the micro contact angle is much larger than the apparent one and approaches to the value of the apparent gradually. 
%
%
%

\begin{figure}
\centering
\includegraphics[scale=.65]{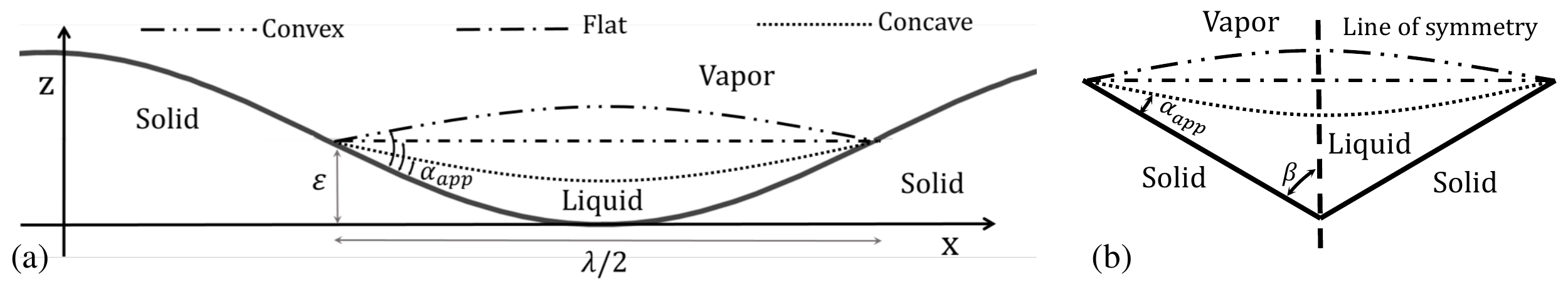}
\caption{Liquid film forming on a wavy surface covering 50\% of the surface (a) surface geometry (b) complementary angle $\beta$}

\label{FIG:wave}
\end{figure}

Surface morphology affects the phase change rate from a liquid film mainly due to its effect on film thickness variation.  The morphology of the substrate may also alter the convex or concave nature of the film surface and change the direction of pressure jump. The phase change rate is a function of the film thickness and the pressure jump across the vapor-liquid interface, given by the augmented Young-Laplace.  The presence of peaks and valleys on the surface has a decisive effect on the film thickness and pressure jump, when the surface features are of the order of nanometeres.  Due to the small amplitude to wavelength ratio of the surface roughness to the separation between two surface peaks in manufactured surfaces, the surface shape can be modeled by a harmonic wave with an amplitude of $\epsilon$, the average height of surface features, and the wavelength of mean peak spacing, $\lambda$.  
\begin{equation}
\label{eq:wave}
\delta_{wave}=2\epsilon \cos \Bigl( \frac{2\pi s}{\lambda}\Bigr),
\end{equation}
In this representation $s$ is the axis along the mean surface and $\delta_{wave}$ is the distance of the surface from the axis (\autoref{FIG:wave}).
The liquid covers the surface only partially, forming droplets, the shapes of which are strong functions of the apparent contact angle.  The difference between the apparent contact angle, $\alpha$, and the angle the line of symmetry makes with the surface, $\beta$, (\autoref{FIG:wave} (b)), can result in nominally concave, flat or convex surfaces away from the close neighborhood of the solid surface, having a profound effect on the rate of phase change.

\begin{figure*}
\centering
\includegraphics[scale=0.8,trim={0.2cm 4.5cm 0.2 2cm},clip]{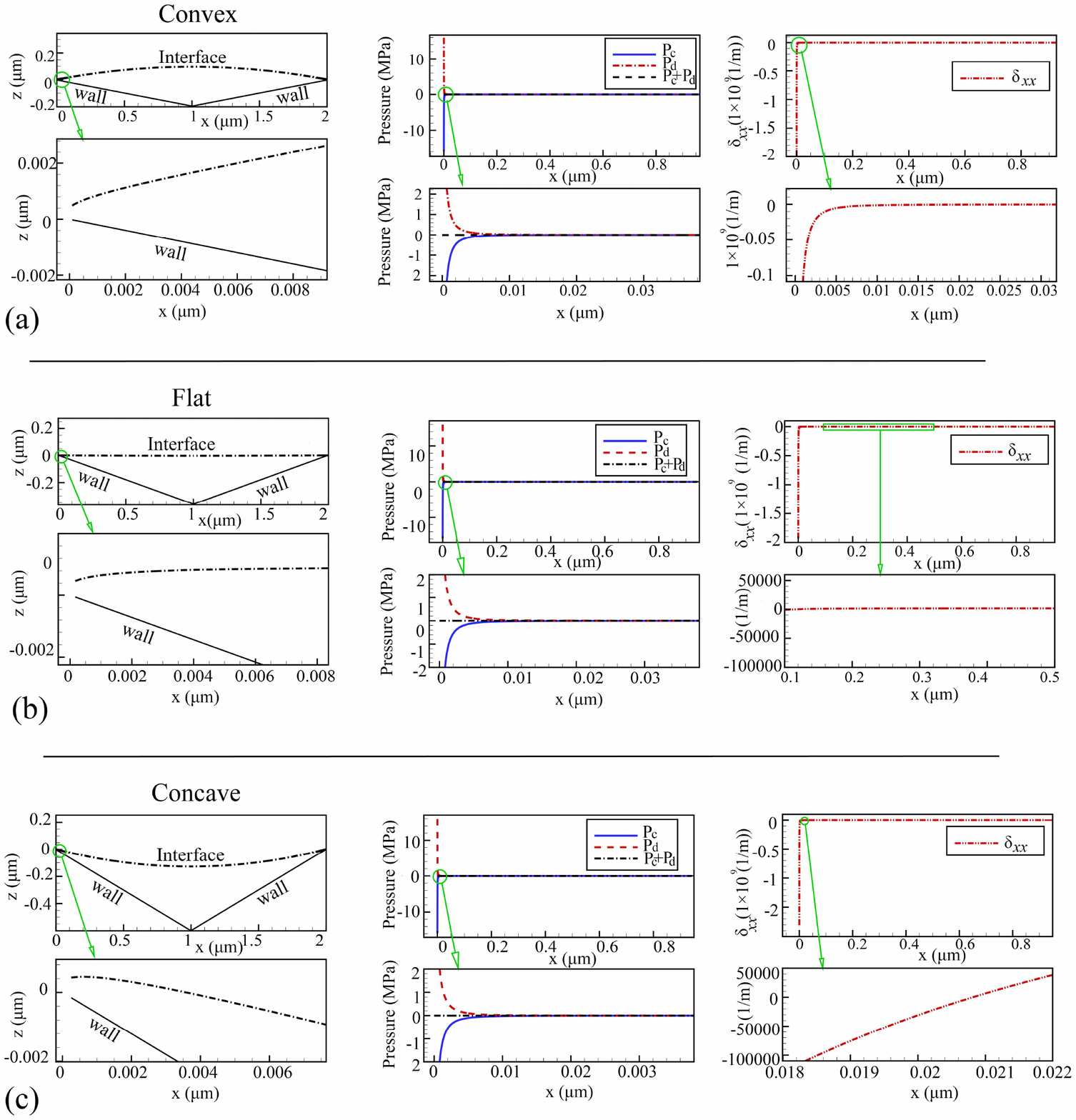}
\caption{Droplet shape, pressures and the second derivative of film thickness, $\delta_{xx}$ on V-shaped, micro structured  (a) convex, (b) flat and (c) concave surfaces }
\label{fig:dropFormation}
\end{figure*}
In a V-shaped surface, the concavity or convexity is a function of the apparent contact angle $\alpha_{app}$ and the angle $\beta$ which is the angle that complements the angle between structured substrate and imaginary line of symmetry. If the angle $(\pi/2 -\beta)$ is smaller than the equilibrium apparent contact angle, $\alpha_\mathit{app}$, the liquid pressure will be larger than the vapor and the interface will be nominally convex, and when $(\pi/2 -\beta) < \alpha_\mathit{app}$ the nominal surface will be concave and the liquid pressure less than that of vapor. In the limiting case of the equality of the two angles $\alpha_{app}$ and $(\pi/2 -\beta) $, the surface will be nominally flat with equal pressures on the liquid and vapor sides (\autoref{FIG:wave} (b)).  The qualifier ``nominal" is used to indicate the fact that the surface shape in the close proximity of the wall may not conform to these shapes due to the effect of disjoining pressure, but far away from the solid surface where dispersion effects vanish the above conclusions apply, hence the use of the term ``nominal".
Initially droplet film profiles formed in a V-shaped (i.e. triangular) groove are studied. \autoref{fig:dropFormation} 
(a), (b) and (c) show the results of simulation for convex, flat and concave shaped cylindrical droplets, respectively.  All simulations refer to a system with an $\alpha$\textsubscript{app} of $19.7^\circ$, the width of triangular surface is $2 \mu {\rm m}$ with fluid depths of $0.2$, $0.4$ and $0.6\mu m$ on the line of symmetry.  For each case, droplet shape, a close up view in the vicinity of the contact line, variation of capillary and disjoining pressures and the curvature, $\kappa$ are plotted. In all three cases in the extreme close vicinity of the solid surface---of the order of 100 nano-meters---the disjoining pressure is dominant resulting in a large negative contribution and a curved down film profile, due to the nature of disjoining pressure term and the sign of Hamaker constant.   Away from the contact line when the dispersion effects vanish, the pressure jump across the interface is balanced only by the capillary term, which results is a constant radius of curvature.   For case (a) the surface is convex, away from the substrate, and the film profile is also convex close to the wall, hence the second derivative is always negative for this case.  In the limiting case of a nominally flat profile, case (b), the surface is flat away from the solid wall but convex near it.  Therefore the second derivative is initially negative, approaching zero away from the contact line.  The nominally concave case (c) starts with a negative second derivative at the wall and the surface changes from concave to convex when the disjoining pressure subsides and the second derivative changes sign to positive away from the contact line.

When the phase change heat transfer on an interface is considered, the mass flux due to phase change is given as \cite{Wayner1976},
\begin{equation}
		\label{Eq:evap3}
		m_{e}''=a\,(T_{lv}-T_v)+b\,(P_l-P_v),
\end{equation}
where $a$ and $b$ are constants defined as,		
\begin{equation}
 \label{eq:evap4}
		a=\frac{2c}{2-c} \Big(\frac{M}{2 \pi \Re T_{lv}}\Big)^\frac{1}{2}    \Big(\frac{M P_v h_{lv}}{\Re T_v T_{lv}} \Big)
\end{equation}
\begin{equation}
		\label{eq:evap5}
		b=\frac{2c}{2-c} \Big(\frac{M}{2 \pi \Re T_{lv}}\Big)^\frac{1}{2} \Big(\frac{P_v V_l}{\Re T_v T_{lv}} \Big)
		\end{equation}
where $c$ is the accommodation coefficient, $h_{lv}$ is the latent heat of evaporation, $M$ the molecular weight, $\Re$ universal gas constant, $V$ molar volume of liquid phase, $T_v$ vapor temperature, and $T_{lv}$ liquid vapor interface temperature.  \autoref{Eq:evap3} shows that the pressure jump at the interface affects the direction of mass flux, that is whether the process will be one of evaporation or condensation.  Although the dominant driving force is the temperature difference between the interface and the vapor, in the case of an isothermal process, the interface pressure jump becomes the only driving potential of phase change.  Therefore, extending the findings presented in \autoref{fig:dropFormation}, in addition to the effect of other surface characteristics, on a surface with surface asperities, the magnitude of asperities and the separation between them may contribute to evaporation or condensation on the same surface simultaneously, due to the effect of the surface shape.

\begin{figure*}
\centering
\includegraphics[scale=.32]{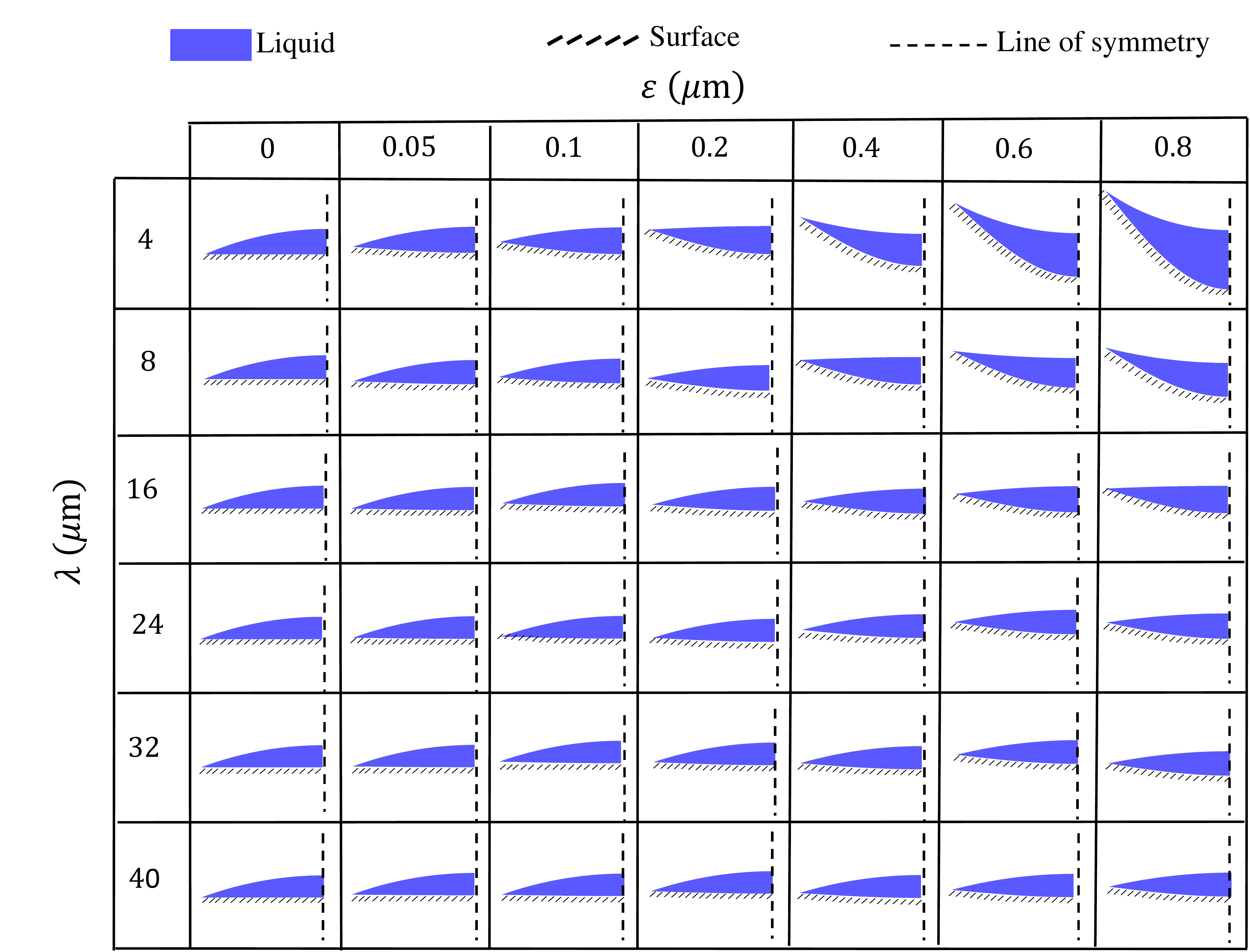}
\caption{Droplet shapes with $\alpha_{app}=20^\circ$ }
\label{FIG:lambda4}
\end{figure*}

\autoref{FIG:wave} shows a cosine wave that can be used to simulate a non-flat concave surface formed due to the presence of asperities.  This approximation can be justified due to the fact that the surface roughness to separation between surface peaks ratio, $2\epsilon/\lambda$, varies between $1/10$\,--$1/100$, with a typical value of $1/20$.    Except for marginal cases, convex surfaces will results in the formation of a convex film, for which the pressure jump will be in the direction of evaporation and not condensation mass transfer.  \autoref{FIG:lambda4} shows various liquid shapes for generated surface profiles with different $2\epsilon/\lambda$ ratios filled 50\% by the liquid.
 In this simulation the separation between two roughness peaks, $\lambda$ is $4 \mu{\rm m}$ with an apparent contact angle, $\alpha_{app}=20^\circ$ \textcolor{red}.  The first case is with a roughness peak, $\epsilon =0 \mu{\rm m}$ which corresponds to a absolutely flat surface.  The next three are cases, $\epsilon =0.05,\, 1$ and $2 \mu{\rm m}$, where the liquid film formed is convex, but increasingly flatter with increased radius of curvature.  The last three cases for $\epsilon =4,\, 6$ and $8 \mu{\rm m}$ are concave.  It should be noted that these particular values of surface peaks are used here to demonstrate the effect, although some of the large $\epsilon$ value do not correspond to any manufactured surfaces seen in common applications.
%
		
\begin{figure}
\centering
\includegraphics[scale=0.9]{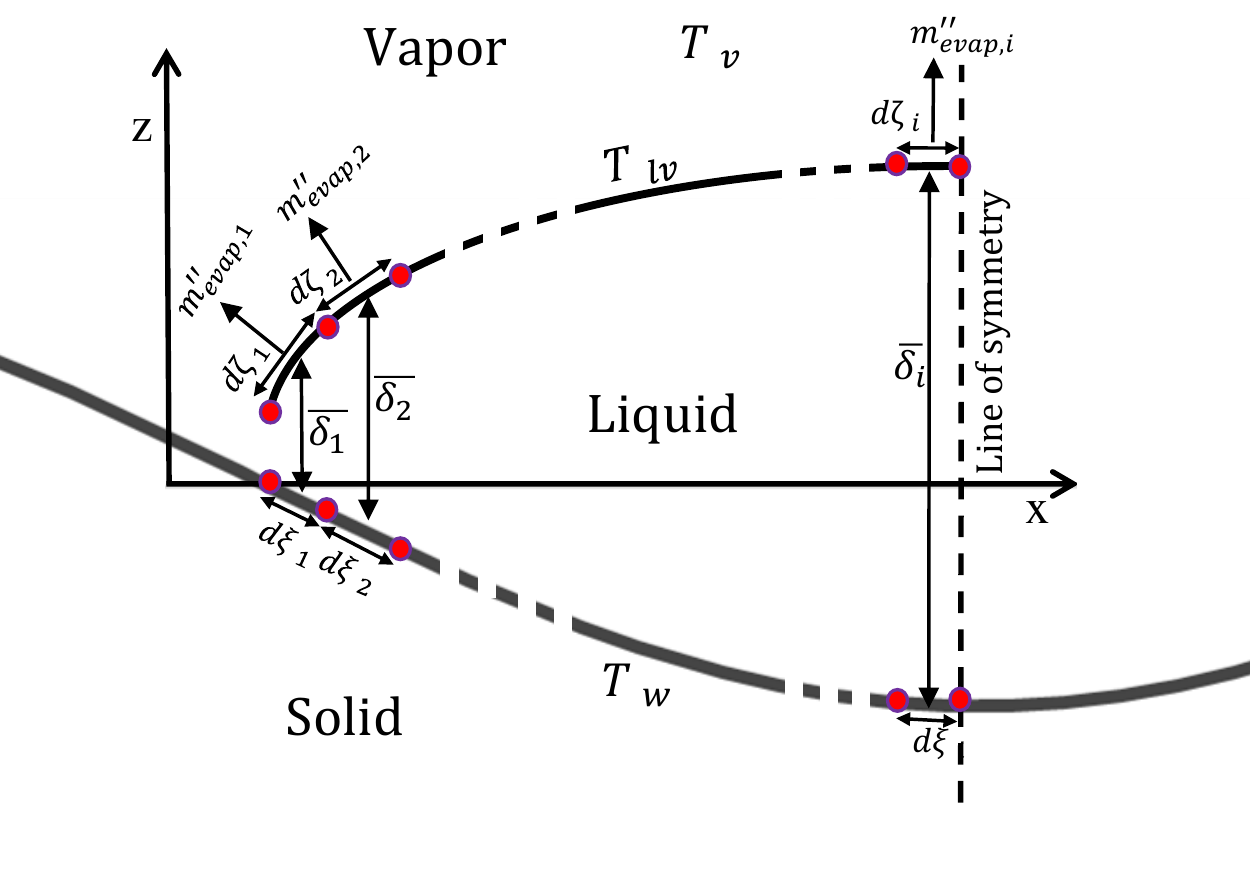}
\caption{Domain discretization for mass flux integration. }
\label{FIG:evap_scheme}
\end{figure}

Due to the very thin liquid films, one can assume a one-dimensional heat conduction in the liquid, which, when coupled to \autoref{Eq:evap3}, results in an expression for mass flux in terms of the film thickness $\delta$ and the solid wall temperature rather than the liquid-vapor interface temperature \cite{Moosman1980}.  
\begin{equation}
	\label{eq:evap8}
		m_e''= \frac {a(T_{w}-T_v)+b(P_l-P_v)}{1+a \delta h_{l v}/k_l}.
\end{equation}		
Heat flux on the interface can be calculated by using the mass flux as, $q''=m'' h_{l v}$, which when integrated over the entire interface of the film will give the rate of total mass transfer and hence heat transfer.   \autoref{FIG:evap_scheme} shows the domain of integration for the calculation of mass flux.  Integration over the interface and the wall gives,
\begin{equation}
\label{eq:evap9}
m'_{evap}= \int_\mathit{interface} \big( a\,( T_{w}-T_{l v}) +b\,(P_v-P_l) \big) \ d\zeta= \int_\mathit{wall}k_l  \frac{T_w-T_{l v}}{\delta h_{l v}} \ d\xi ,
\end{equation} 
and eliminating $T_{l v}$ in \autoref{eq:evap9},
\begin{equation}
		\label{eq:evap11}
		m_{evap}'= \int_{0}^{S_{max}} \frac {a(T_{w}-T_v)+b(P_l-P_v)}{1+ (a\delta h_{lv}/k_l) (d \xi / d \zeta) } d\zeta,
\end{equation}
%
%
%
where $S_{max}$ is the interface length on the surface of a droplet.  This integration is numerically carried out using the discretization depicted in \autoref{FIG:evap_scheme}.
%
%

%
%
%

A literature review revealed that for  materials 
commercially used in heat transfer devices with smooth surface finish, the values of average surface roughness varies between 0.05 to 0.8$\ \mu {\rm m}$ and the 2 to 40$\ \mu{\rm m}$ crest spacing.  
A set of simulations were done for ammonia, which has $19.7^\circ$ apparent contact angle, a fluid commonly used in heat pipes, for which thermophysical properties given in \cite{Stephan1992} are used. The surfaces are covered by $50\%$ with the fluid.  Due to the acute contact angle of ammonia, film profiles vary from convex to concave with increasing magnitude surface peaks. In case of peaks with a higher frequency of occurrence, the average film thickness is smaller which results in a higher rate of evaporation or condensation mass transfer and as a result higher heat flux.  Initially the surface film profile is calculated, the phase change rate calculations are done {\it a posteriori} assuming a quasi-steady system with a temperature difference $\Delta T=\pm 1{\rm C^\circ}$, where a positive value corresponds to evaporation and negative to condensation.  Isothermal, case is also studied, where the phase change is only due to the pressure jump at the interface, type governed by the sign of the pressure jump, hence by the convexity or concavity of the nominal interface.
\autoref{tab:isothermal} show the results for the isothermal problem.
\begin{table}
\caption{\label{tab:isothermal} Phase change heat flux for isothermal cases, ${\rm (10^{-4} \times kW/m^2)}$ }
\begin{tabular}{p{0.8cm}p{0.4cm}|p{1.35cm}p{1.35cm}p{1.35cm}p{1.35cm}p{1.35cm}p{1.35cm}p{1.35cm}  }
&&&&&$\varepsilon (\mu m)$\\
&   & 0 & 0.05 & 0.1 & 0.2 & 0.4 & 0.6 & 0.8\\
   \hline
&4  & 168.91	&138.33	&104.17	&27.48	&\textbf{-133.43}	& \textbf{-219.83} &\textbf{-260.64}\\
&8  &48.47	&43.97	&39.22	&28.97	&6.95	&\textbf{-16.30}	&\textbf{-34.51}\\
$\lambda(\mu m)$&16 & 13.69	&13.05	&12.37	&10.98	&8.05	&4.97	&1.83\\
&24 &6.49	&6.28	&6.08	&5.65	&4.74	&3.79	&2.82\\
&32 &3.82	&3.73	&3.63	&3.44	&3.05	&2.64	&2.22\\
&40 &2.52	&2.47	&2.43	&2.33	&2.12	&1.91	&1.69\\
\end{tabular}
\end{table}
In this case the heat transfer occurs only due to the pressure difference given in \autoref{Eq:evap3}.  Due to the geometry, only 5 cases have a nominally concave film profile, whereas the remaining 37 are convex.  In case of the concave surfaces the mass transfer is from the liquid to the vapor, i.e. evaporation and the opposite is valid for the nominally convex surfaces.  The magnitudes of phase change rate are  $\mathcal{O}( 10^{-4}$ -- $10^{-2})\, {\rm kW/m^2}$, approximately 2 to 4 orders of magnitude less than those due to a $1^\circ{\rm{ C}}$ difference given in \autoref{tab:evaporation} and \autoref{tab:condensation}.

This stark contrast would reduce linearly with reduced temperature differences, for example for a temperature difference of $\Delta T=\pm 0.01{\rm C^\circ}$ the contribution of pressure jump and temperature difference would be comparable for the small wavelength cases. 
\autoref{tab:isot2} shows a result  for a temperature difference of $\Delta T=\pm 0.01{\rm C^\circ}$ only 4 $\mu m \, \lambda$ case.
\begin{table}
\caption{\label{tab:isot2} Phase change heat flux for $\Delta T=0.01\rm K$ and $\lambda=4\mu {\rm m}$, ${\rm (10^{-2} \times kW/m^2)}$}
\begin{tabular}{p{2.5cm}|p{1.2cm}p{1.2cm}p{1.2cm}p{1.2cm}p{1.2cm}p{1.2cm}p{0.9cm}  }
&&&&&$\varepsilon (\mu m)$\\
   & 0 & 0.05 & 0.1 & 0.2 & 0.4 & 0.6& 0.8\\
          \hline
Evaporation &6.69	&6.53	&6.30	&5.74	&\textbf{4.49}	&\textbf{3.39}	&\textbf{2.64}
 \\
Condensation&3.31	&3.76	&4.22	&5.19	&\textbf{7.16}	&\textbf{7.78}	&\textbf{7.85}
\\
Isothermal& 1.69	&1.38	&1.04	&0.27	&\textbf{1.33}	&\textbf{2.20}	&\textbf{2.61}
\\
\end{tabular}
\end{table}
The phase change rate is at a maximum for the smallest separation distances, $\lambda$.  This is due to the small radii of curvature encountered in these cases which result in larger pressure jumps. In \autoref{tab:evaporation} the phase change rate results of simulations with a positive temperature difference, i.e. evaporation, are given.

\begin{table}
\caption{\label{tab:evaporation} Evaporation heat flux for $\Delta T=1\rm K$, ${\rm (kW/m^2)}$}
\begin{tabular}{p{0.8cm}p{0.4cm}|p{1.4cm}p{1.4cm}p{1.4cm}p{1.4cm}p{1.4cm}p{1.4cm}p{0.9cm}  }
&&&&&$\varepsilon (\mu m)$\\
&   & 0 & 0.05 & 0.1 & 0.2 & 0.4 & 0.6 & 0.8\\
          \hline
&4& 5.019 & 5.159 & 5.272 & 5.471  &\textbf{ 5.814}  & \textbf{5.562}  & \textbf{5.215} \\

&8& 3.051 & 3.091& 3.126 & 3.186  & 3.258  & \textbf{3.306}  & \textbf{3.303}  \\
$\lambda(\mu m)$&16 & 1.791 & 1.801 & 1.812& 1.831  & 1.857 & 1.876 & 1.885 \\
&24& 1.295 & 1.300 & 1.305 & 1.314  & 1.331 & 1.340  & 1.348  \\
&32& 1.024 & 1.027& 1.029 & 1.035  & 1.045  & 1.053  & 1.059  \\
&40& 0.851 & 0.853 & 0.855 & 0.859  & 0.866  & 0.873  & 0.878 
\end{tabular}
\end{table}
The results show that the evaporation rates are at a maximum when the surface covered by the film is at a minimum.  Furthermore, increasing the amplitude of surface features enhances the heat transfer reaching a maximum at {\it ca.} $0.4\ \mu{\rm m}$, decreasing with increasing asperity size.  This behavior is due to the fact that, starting from a flat surface, where $\varepsilon=0$, introduction of asperities increases the evaporation rate due to the generation of a thinner film, bringing the interface to a closer proximity of the isothermal wall.  Although this increases the radius of curvature---an unfavorable effect to evaporation---it is nevertheless outweighed by the effect of a smaller film thickness given in \autoref{eq:evap8}. Further increase in the asperity amplitude changes the shape of the surface from convex to concave, this time hindering evaporation.  All this results in a maximum evaporation rate around  $\varepsilon=0.4\ \mu{\rm m}$.  Similar behavior is observed for separation distances $8\ \mu{\rm m}<\lambda<40\ \mu{\rm m}$, but in these cases the radius of curvature is much larger, hence the effect less pronounced and even exists beyond the range where simulations were done.  The results for the condensation flux are given in \autoref{tab:condensation}, a behavior identical to the evaporation case can be observed due to similar physical phenomena. 
\begin{table}
\caption{\label{tab:condensation} Condensation heat flux for $\Delta T=1\rm K$, ${\rm (kW/m^2)}$}
\begin{tabular}{p{0.8cm}p{0.4cm}|p{1.4cm}p{1.4cm}p{1.4cm}p{1.4cm}p{1.4cm}p{1.4cm}p{0.9cm}  }
&&&&&$\varepsilon (\mu m)$\\
&   & 0 & 0.05 & 0.1 & 0.2 & 0.4 & 0.6 & 0.8\\
          \hline
&4 & 4.985& 5.131& 5.251& 5.465  & \textbf{5.841} & \textbf{5.606}  & \textbf{5.267} \\
&8&3.0421&	3.082&	3.118&	3.180&	3.256&\textbf{3.310}&\textbf{3.311}\\
$\lambda(\mu m)$&16 &1.788	&1.799&	1.810&	1.829&	1.856&1.875	&1.885\\
&24 & 1.294&	1.298&	1.304&	1.313&	1.330&1.339&	1.348\\
&32 &  1.023&	1.026&	1.029&	1.035&	1.045&1.053&	1.058\\
&40 &0.851&	0.853&	0.855&	0.858&	0.866&0.872&	0.878\\
\end{tabular}
\end{table}
 It can be concluded that to enhance the rate of evaporation or condensation, when this is desired, wavy surfaces with short amplitudes are preferable to ones with longer amplitudes, as an example the heat and mass transfer rate increases by a factor of 5 on a $50\%$ wetted surface with $4\ \mu{\rm m} $ compared to a surface with a peak to peak separation $\lambda=32\ \mu{\rm m}$.
\section{Conclusion}
The influence of surface characteristics on phase change rate for thin films (cylindrical droplets), micro meniscus film and liquid filled valleys are studied. The augmented Young-Laplace equation is solved numerically to account for the disperison effects and a phase change model is applied to the resulting fluid interface.  Disjoining pressure with only the dispersion term is considered for partial wetting droplets with static apparent contact angles.  A phase change model is applied and rate of phase change heat transfer from the film is calculated.  The results are shown qualitatively and quantitatively for different temperature difference between surface and vapor. Two important components of phase change rate, pressure and temperature jump contribution have been studied. The effect of temperature jump showed that droplets on peaks (convex droplets) tend to evaporate and condensation would occur simultaneously on valleys for the case which contact angle is more than critical angle (concave droplets).   Due to the low amplitude to wavelength ratio of commonly encountered machined surfaces, surfaces form complementary angles  ($\beta$, \autoref{FIG:wave} (b)) close to $90^\circ$, which along with the apparent contact angle decides on the concavity, flatness or convexity of the resulting 2-D droplet.  In addition to a large complementary angle, when the apparent contact angle is small the film is convex, a rare occurrence.  Even for the smallest droplets the dominant driving potential for phase change is the temperature difference, with the pressure jump having a minor, negligible contribution, except for the isothermal cases where the pressure jump is the only driving potential off phase change and heat transfer.  The major effect of the surface morphology, however is the reduction of film thickness which substantially enhances heat transfer.  The disjoining pressure is particularly effective both in the film shape and heat transfer in the close proximity of the contact line, where the film is the thinnest and the corresponding heat flux a maximum.  In the case of the smallest 2-D droplets the disjoining pressure being dominant, the droplet surface never attains the apparent contact angle observed in macroscopic measurements eliminating the possibility of using this value as a boundary condition.  The current study offers a remedy to this by using  the micro scale contact angle obtained from MD simulations as the boundary condition.
\section{Acknowledgment}
The authors acknowledge Dr.~Yi\u git Akku\c s for his careful review of the manuscript.  This study is financially supported by the Turkish Scientific and Technical Research Council, under grant No. 213M351. 
\bibliographystyle{unsrt}
\bibliography{MyCollection}
\medskip
\end{document}